\documentclass[aps,prd,reprint, superscriptaddress,preprintnumbers, longbibliography, amsmath,amssymb,amsfonts, nofootinbib]{revtex4-2}
\usepackage{graphicx}
\usepackage[dvipsnames]{xcolor}
\usepackage{hyperref}
\usepackage{xurl}
\usepackage{xspace}
\usepackage{ifdraft}
\usepackage{epstopdf}
\usepackage[bottom]{footmisc}
\usepackage{slashed}
\usepackage{xcolor}
\hypersetup{
    colorlinks,
    linkcolor={red!50!black},
    citecolor={blue!50!black},
    urlcolor={red!50!black}
}
\usepackage{tikz-feynman} 

\tikzfeynmanset{compat=1.1.0}
\tikzfeynmanset{/tikzfeynman/warn luatex = false}

\definecolor{jaxoblue}{HTML}{0086FF}

\definecolor{hgreen}{RGB}{25,176,146}
\definecolor{hblue}{RGB}{52,152,219}
\definecolor{hred}{RGB}{216,83,117}
\definecolor{cutred}{RGB}{219,56,49}
\definecolor{hgrey}{RGB}{125,125,125}
\definecolor{darkgreen}{RGB}{59,126,108}
\definecolor{cutred}{RGB}{219,56,49}

\newcommand{\halfLadder}[1]{ {
\begin{tikzpicture}[baseline=7.5pt]
\begin{feynman}
\vertex (mid1) at (-2*0.58, 0);
\vertex (mid2) at (-1*0.58, 0);
\vertex (mid3) at (1*0.58, 0);
\vertex [dot, scale=0.4](d1) at (-0.4*0.58, .5*1.3*0.58){};
\vertex [dot, scale=0.4](d2) at (0*0.58, .5*1.3*0.58){};
\vertex [dot, scale=0.4](d3) at (.4*0.58, .5*1.3*0.58){};
\vertex (a1) at (-3*0.58, 0) {$i$};
\vertex (a2) at (-2*0.58, 1*1.3*0.58) {$\tau_2$};
\vertex (a3) at (-1*0.58, 1*1.3*0.58) {$\tau_3$};
\vertex (a4) at (1*0.58, 1*1.3*0.58) {$\tau_{n-1}$};
\vertex (a5) at (2*0.58, 0) {$j$};
\diagram{
(mid1)--[thick](a1),
(mid1)--[thick,gluon](a2),
(mid1)--[thick](mid2),
(mid3)--[thick](mid2),
(mid2)--[thick,gluon](a3),
(mid3)--[thick,gluon](a4),
(mid3)--[thick](a5),
};
\end{feynman}
\end{tikzpicture}
}
}

\newcommand{\diskEightpointWithDot}[8]{ {
\begin{tikzpicture}[baseline=-2pt,squarednode/.style={rectangle, draw=black, fill=black!5, very thick, minimum size=5mm}]
\begin{feynman}
\vertex(mid1) at (0,0) {};
\draw (mid1) circle(0.51);
\vertex[dot,hred] (a5) at (-.5,0) {};
\vertex[dot,scale=0.9] (a6) at (-0.3535534,-0.3535534) {};
\vertex[dot,scale=0.9] (a7) at  (0,-.5) {};
\vertex[dot,scale=0.9] (a8) at (0.3535534,-0.3535534) {};
\vertex[squarednode,scale=0.25] (a1) at  (.5,0){};
\vertex[dot,scale=0.9] (a2) at (0.3535534,0.3535534) {};
\vertex[dot,hred] (a3) at (0,.5) {};
\vertex[dot,scale=0.9] (a4) at (-0.3535534,0.3535534) {};
\diagram{
(a#1)--[ultra thick](a#2),
(a#3)--[ultra thick](a#4),
(a#5)--[ultra thick](a#6),
(a#7)--[ultra thick](a#8)
};
\end{feynman}
\end{tikzpicture}
}
}

\newcommand{\diskSixteenpointA}[8]{ {
\begin{tikzpicture}[baseline=-2pt,squarednode/.style={rectangle, draw=black, fill=black!5, very thick, minimum size=5mm}]
\begin{feynman}
\vertex(mid1) at (0,0) {};
\draw (mid1) circle(0.5);
\vertex[dot,scale=0.5] (a1) at (-0.490, 0.09754) {};
\vertex[dot,scale=0.5] (a2) at (-0.415, 0.277){};
\vertex[dot,scale=0.5] (a3) at (-0.277, 0.41){};
\vertex[dot,scale=0.5] (a4) at (-0.0975, 0.49){};
\vertex[dot,scale=0.5] (a5) at (0.0975, 0.490){};
\vertex[dot,scale=0.5] (a6) at (0.277, 0.415){};
\vertex[dot,scale=0.5] (a7) at (0.415, 0.27){};
\vertex[dot,scale=0.5] (a8) at (0.490, 0.097){};
\vertex[dot,scale=0.5] (a9) at (0.490, -0.097){};
\vertex[dot,scale=0.5] (a10) at (0.415, -0.277){};
\vertex[dot,scale=0.5] (a11) at (0.277, -0.41){};
\vertex[dot,scale=0.5] (a12) at (0.0975, -0.490){};
\vertex[dot,scale=0.5] (a13) at (-0.0975, -0.490){};
\vertex[dot,scale=0.5] (a14) at (-0.277, -0.415){};
\vertex[dot,scale=0.5] (a15) at (-0.415, -0.277){};
\vertex[dot,scale=0.5] (a16) at (-0.490, -0.097){};
\vertex[dot,scale=0.9] (a17) at (-3, 0);
\diagram{
(a3)--[thick,quarter left](a16),
(a15)--[thick,quarter left](a12),
(a13)--[thick,quarter left](a10),
(a9)--[thick,quarter left](a6),
(a#1)--[thick,quarter left,hred](a#2),
(a#3)--[thick,quarter left,hred](a#4),
(a#5)--[thick,quarter left,hred](a#6),
(a#7)--[thick,quarter left,hred](a#8)
};
\end{feynman}
\end{tikzpicture}
}
}

\newcommand{\diskSixteenpointB}[8]{ {
\begin{tikzpicture}[baseline=-2pt,squarednode/.style={rectangle, draw=black, fill=black!5, very thick, minimum size=5mm}]
\begin{feynman}
\vertex(mid1) at (0,0) {};
\draw (mid1) circle(0.5);
\vertex[dot,scale=0.5] (a1) at (-0.490, 0.09754) {};
\vertex[dot,scale=0.5] (a2) at (-0.415, 0.277){};
\vertex[dot,scale=0.5] (a3) at (-0.277, 0.41){};
\vertex[dot,scale=0.5] (a4) at (-0.0975, 0.49){};
\vertex[dot,scale=0.5] (a5) at (0.0975, 0.490){};
\vertex[dot,scale=0.5] (a6) at (0.277, 0.415){};
\vertex[dot,scale=0.5] (a7) at (0.415, 0.27){};
\vertex[dot,scale=0.5] (a8) at (0.490, 0.097){};
\vertex[dot,scale=0.5] (a9) at (0.490, -0.097){};
\vertex[dot,scale=0.5] (a10) at (0.415, -0.277){};
\vertex[dot,scale=0.5] (a11) at (0.277, -0.41){};
\vertex[dot,scale=0.5] (a12) at (0.0975, -0.490){};
\vertex[dot,scale=0.5] (a13) at (-0.0975, -0.490){};
\vertex[dot,scale=0.5] (a14) at (-0.277, -0.415){};
\vertex[dot,scale=0.5] (a15) at (-0.415, -0.277){};
\vertex[dot,scale=0.5] (a16) at (-0.490, -0.097){};
\vertex[dot,scale=0.9] (a17) at (-3, 0);
\diagram{
(a3)--[thick,quarter left](a16),
(a15)--[thick,quarter left](a12),
(a13)--[thick,quarter left](a10),
(a9)--[thick,quarter left](a6),
(a#1)--[thick,quarter left,hred](a#2),
(a#3)--[thick,quarter left,hred](a#4),
(a#5)--[thick,hred](a#6),
(a#7)--[thick,hred](a#8)
};
\end{feynman}
\end{tikzpicture}
}
}

\newcommand{\diskSixteenpointC}[8]{ {
\begin{tikzpicture}[baseline=-2pt,squarednode/.style={rectangle, draw=black, fill=black!5, very thick, minimum size=5mm}]
\begin{feynman}
\vertex(mid1) at (0,0) {};
\draw (mid1) circle(0.5);
\vertex[dot,scale=0.5] (a1) at (-0.490, 0.09754) {};
\vertex[dot,scale=0.5] (a2) at (-0.415, 0.277){};
\vertex[dot,scale=0.5] (a3) at (-0.277, 0.41){};
\vertex[dot,scale=0.5] (a4) at (-0.0975, 0.49){};
\vertex[dot,scale=0.5] (a5) at (0.0975, 0.490){};
\vertex[dot,scale=0.5] (a6) at (0.277, 0.415){};
\vertex[dot,scale=0.5] (a7) at (0.415, 0.27){};
\vertex[dot,scale=0.5] (a8) at (0.490, 0.097){};
\vertex[dot,scale=0.5] (a9) at (0.490, -0.097){};
\vertex[dot,scale=0.5] (a10) at (0.415, -0.277){};
\vertex[dot,scale=0.5] (a11) at (0.277, -0.41){};
\vertex[dot,scale=0.5] (a12) at (0.0975, -0.490){};
\vertex[dot,scale=0.5] (a13) at (-0.0975, -0.490){};
\vertex[dot,scale=0.5] (a14) at (-0.277, -0.415){};
\vertex[dot,scale=0.5] (a15) at (-0.415, -0.277){};
\vertex[dot,scale=0.5] (a16) at (-0.490, -0.097){};
\vertex (a17) at (-3, 0);
\diagram{
(a3)--[thick,quarter left](a16),
(a15)--[thick,quarter left](a12),
(a13)--[thick,quarter left](a10),
(a9)--[thick,quarter left](a6),
(a#1)--[thick,quarter left,hred](a#2),
(a#3)--[thick,quarter left,hred](a#4),
(a#5)--[thick,quarter left,hred](a#6),
(a#7)--[thick,hred](a#8)
};
\end{feynman}
\end{tikzpicture}
}
}

\newcommand{\diskEightpoint}[8]{ {
\begin{tikzpicture}[baseline=-2pt,squarednode/.style={rectangle, draw=black, fill=black!5, very thick, minimum size=5mm}]
\begin{feynman}
\vertex(mid1) at (0,0) {};
\draw (mid1) circle(0.51);
\vertex[dot,scale=0.9] (a5) at (-.5,0) {};
\vertex[dot,scale=0.9] (a6) at (-0.3535534,-0.3535534) {};
\vertex[dot,scale=0.9] (a7) at  (0,-.5) {};
\vertex[dot,scale=0.9] (a8) at (0.3535534,-0.3535534) {};
\vertex[squarednode,scale=0.25] (a1) at  (.5,0){};
\vertex[dot,scale=0.9] (a2) at (0.3535534,0.3535534) {};
\vertex[dot,scale=0.9] (a3) at (0,.5) {};
\vertex[dot,scale=0.9] (a4) at (-0.3535534,0.3535534) {};
\diagram{
(a#1)--[ultra thick](a#2),
(a#3)--[ultra thick, hred](a#4),
(a#5)--[ultra thick, hred](a#6),
(a#7)--[ultra thick,  hred](a#8)
};
\end{feynman}
\end{tikzpicture}
}
}

\newcommand{\diskEightpointhABig}[8]{ {
\begin{tikzpicture}[baseline=-2pt,squarednode/.style={rectangle, draw=black, fill=black!5, very thick, minimum size=5mm},scale=1.02]
\begin{feynman}
\vertex(mid1) at (0,0) {};
\draw (mid1) circle(0.62);
\vertex[dot,scale=0.9] (a5) at (-.5*1.2,0) {};
\vertex[dot,scale=0.9] (a6) at (-0.3535534*1.2,-0.3535534*1.2) {};
\vertex[dot,scale=0.9] (a7) at  (0*1.2,-.5*1.2) {};
\vertex[dot,scale=0.9] (a8) at (0.3535534*1.2,-0.3535534*1.2) {};
\vertex[squarednode,scale=0.25] (a1) at  (.5*1.2,0*1.2){};
\vertex[dot,scale=0.9] (a2) at (0.3535534*1.2,0.3535534*1.2) {};
\vertex[dot,scale=0.9] (a3) at (0*1.2,.5*1.2) {};
\vertex[dot,scale=0.9] (a4) at (-0.3535534*1.2,0.3535534*1.2) {};
\diagram{
(a#1)--[ultra thick](a#2),
(a#5)--[ultra thick](a#6),
(a#3)--[ultra thick,hred](a#4),
(a#7)--[ultra thick,hred](a#8)
};
\end{feynman}
\end{tikzpicture}
}
}

\newcommand{\diskEightpointhAB}[8]{ {
\begin{tikzpicture}[baseline=-2pt,squarednode/.style={rectangle, draw=black, fill=black!5, very thick, minimum size=5mm}]
\begin{feynman}
\vertex(mid1) at (0,0) {};
\draw (mid1) circle(0.51);
\vertex[dot,scale=0.9] (a5) at (-.5,0) {};
\vertex[dot,scale=0.9] (a6) at (-0.3535534,-0.3535534) {};
\vertex[dot,scale=0.9] (a7) at  (0,-.5) {};
\vertex[dot,scale=0.9] (a8) at (0.3535534,-0.3535534) {};
\vertex[squarednode,scale=0.25] (a1) at  (.5,0){};
\vertex[dot,scale=0.9] (a2) at (0.3535534,0.3535534) {};
\vertex[dot,scale=0.9] (a3) at (0,.5) {};
\vertex[dot,scale=0.9] (a4) at (-0.3535534,0.3535534) {};
\diagram{
(a#1)--[ultra thick](a#2),
(a#5)--[ultra thick](a#6),
(a#3)--[ultra thick,hred](a#4),
(a#7)--[ultra thick,hred](a#8)
};
\end{feynman}
\end{tikzpicture}
}
}

\newcommand{\diskEightpointAB}[8]{ {
\begin{tikzpicture}[baseline=-2pt,squarednode/.style={rectangle, draw=black, fill=black!5, very thick, minimum size=5mm}]
\begin{feynman}
\vertex(mid1) at (0,0) {};
\draw (mid1) circle(0.51);
\vertex[dot,scale=0.9] (a5) at (-.5,0) {};
\vertex[dot,scale=0.9] (a6) at (-0.3535534,-0.3535534) {};
\vertex[dot,scale=0.9] (a7) at  (0,-.5) {};
\vertex[dot,scale=0.9] (a8) at (0.3535534,-0.3535534) {};
\vertex[squarednode,scale=0.25] (a1) at  (.5,0){};
\vertex[dot,scale=0.9] (a2) at (0.3535534,0.3535534) {};
\vertex[dot,scale=0.9] (a3) at (0,.5) {};
\vertex[dot,scale=0.9] (a4) at (-0.3535534,0.3535534) {};
\diagram{
(a#1)--[ultra thick](a#2),
(a#3)--[ultra thick](a#4),
(a#5)--[ultra thick](a#6),
(a#7)--[ultra thick](a#8)
};
\end{feynman}
\end{tikzpicture}
}
}

\newcommand{\diskEightpointABcurve}[8]{ {
\begin{tikzpicture}[baseline=-2pt,squarednode/.style={rectangle, draw=black, fill=black!5, very thick, minimum size=5mm}]
\begin{feynman}
\vertex(mid1) at (0,0) {};
\draw (mid1) circle(0.51);
\vertex[dot,scale=0.9] (a5) at (-.5,0) {};
\vertex[dot,scale=0.9] (a6) at (-0.3535534,-0.3535534) {};
\vertex[dot,scale=0.9] (a7) at  (0,-.5) {};
\vertex[dot,scale=0.9] (a8) at (0.3535534,-0.3535534) {};
\vertex[squarednode,scale=0.25] (a1) at  (.5,0){};
\vertex[dot,scale=0.9] (a2) at (0.3535534,0.3535534) {};
\vertex[dot,scale=0.9] (a3) at (0,.5) {};
\vertex[dot,scale=0.9] (a4) at (-0.3535534,0.3535534) {};
\diagram{
(a#1)--[ultra thick,quarter left](a#2),
(a#3)--[ultra thick,quarter left,looseness=0.7](a#4),
(a#5)--[ultra thick](a#6),
(a#7)--[ultra thick](a#8)
};
\end{feynman}
\end{tikzpicture}
}
}

\newcommand{\diskFourpoint}[4]{ {
\begin{tikzpicture}[baseline=-2pt,squarednode/.style={rectangle, draw=black, fill=black!5, very thick, minimum size=5mm}]
\begin{feynman}
\vertex(mid1) at (0,0) {};
\draw (mid1) circle(0.51);
\vertex[squarednode,scale=0.25] (a1) at (.5,0) {};
\vertex[dot,scale=0.9] (a2) at (0,.5) {};
\vertex[dot,scale=0.9] (a3) at (-0.5,0){};
\vertex[dot,scale=0.9] (a4) at (0,-.5) {};
\diagram{
(a#1)--[ultra thick](a#2),
(a#3)--[ultra thick](a#4),
};
\end{feynman}
\end{tikzpicture}
}
}

\newcommand{\diskSixpointStraight}[6]{ {
\begin{tikzpicture}[baseline=-2pt,squarednode/.style={rectangle, draw=black, fill=black!5, very thick, minimum size=5mm}]
\begin{feynman}
\vertex(mid1) at (0,0) {};
\draw (mid1) circle(0.51);
\vertex[dot,scale=0.9] (a4) at (-.5,0) {};
\vertex[dot,scale=0.9] (a3) at (-0.25,0.4330127) {};
\vertex[dot,scale=0.9] (a2) at (0.25,0.4330127) {};
\vertex[squarednode,scale=0.25] (a1) at (.5,0) {};
\vertex[dot,scale=0.9] (a6) at  (0.25,-0.4330127) {};
\vertex[dot,scale=0.9] (a5) at (-0.25,-0.4330127){};
\diagram{
(a#1)--[ultra thick](a#2),
(a#3)--[ultra thick](a#4),
(a#5)--[ultra thick](a#6),
};
\end{feynman}
\end{tikzpicture}
}
}

\newcommand{\diskSixpointPol}[6]{ {
\begin{tikzpicture}[baseline=-2pt,squarednode/.style={rectangle, draw=black, fill=black!5, very thick, minimum size=5mm}]
\begin{feynman}
\vertex(mid1) at (0,0) {};
\draw (mid1) circle(0.51);
\vertex[dot,scale=0.9] (a4) at (-.5,0) {};
\vertex[dot,hred] (a3) at (-0.25,0.4330127) {};
\vertex[dot,scale=0.9] (a2) at (0.25,0.4330127) {};
\vertex[squarednode,scale=0.25] (a1) at (.5,0) {};
\vertex[dot,scale=0.9] (a6) at  (0.25,-0.4330127) {};
\vertex[dot,scale=0.9] (a5) at (-0.25,-0.4330127){};
\diagram{
(a#1)--[ultra thick](a#2),
(a#3)--[ultra thick](a#4),
(a#5)--[ultra thick](a#6),
};
\end{feynman}
\end{tikzpicture}
}
}

\newcommand{\crustlessSet}[0]{ {
\begin{tikzpicture}[baseline=-2.5pt]
\begin{feynman}
\vertex (mid1) at (0,0) {};
\draw[thick]  (mid1) circle(0.5*0.52);
\vertex (a4) at (0*0.52,-.5*0.52) ;
\vertex (a3) at (0.4330127*0.52,-0.25*0.52) ;
\vertex (a2) at (0.4330127*0.52,0.25*0.52) ;
\vertex (a1) at (0*0.52,.5*0.52) ;
\vertex (a6) at  (-0.4330127*0.52,0.25*0.52) ;
\vertex (a5) at (-0.4330127*0.52,-0.25*0.52);
\diagram{
(a1)--[thick](a4),
(a2)--[thick](a5),
(a3)--[thick](a6)
};
\end{feynman}
\end{tikzpicture}
}
}

\newcommand{\crustlessSetinLine}[0]{ {
\begin{tikzpicture}[baseline=-2.5pt]
\begin{feynman}
\vertex (mid1) at (0,0) {};
\draw[thick]  (mid1) circle(0.5*0.35);
\vertex (a4) at (0*0.35,-.5*0.35) ;
\vertex (a3) at (0.4330127*0.35,-0.25*0.35) ;
\vertex (a2) at (0.4330127*0.35,0.25*0.35) ;
\vertex (a1) at (0*0.35,0.5*0.35) ;
\vertex (a6) at  (-0.4330127*0.35,0.25*0.35) ;
\vertex (a5) at (-0.4330127*0.35,-0.25*0.35);
\diagram{
(a1)--[thick](a4),
(a2)--[thick](a5),
(a3)--[thick](a6)
};
\end{feynman}
\end{tikzpicture}
}
}

\newcommand{\eightPointCut}[1]{ {
\begin{tikzpicture}[baseline=-2pt,squarednode/.style={rectangle, draw=black, fill=black!5, very thick, minimum size=5mm}]
\begin{feynman}
\vertex(mid1) at (-1.2,0) {};
\vertex(mid3) at (0,0) {};
\vertex(mid5) at (.92*1.2,0) {};
\draw (mid1) circle(0.6);
\draw (mid3) circle(0.6);
\draw (mid5) circle(0.5);
\vertex[dot,scale=0.9] (a1) at (-1*1.2,0.5*1.2) {};
\vertex[dot,scale=0.9] (a2) at (-1*1.2,-0.5*1.2) {};
\vertex[dot,scale=0.9] (a3) at (-1.35*1.2,0.36*1.2) {};
\vertex[dot,scale=0.9] (a4) at  (-1.35*1.2,-0.36*1.2) {};
\vertex[dot,scale=0.9] (a5) at (0*1.2,0.5*1.2) {};
\vertex[dot,scale=0.9] (a6) at (0*1.2,-0.5*1.2){};
\vertex[dot,scale=0.9] (a7) at (1*1.2,.4*1.2){};
\vertex[squarednode,scale=0.25] (a8) at (1*1.2,-.4*1.2){};
\vertex[dot,scale=0.9] (a9) at (-0.5*1.2,0*1.2){};
\vertex[dot,scale=0.9] (a10) at (0.5*1.2,0*1.2){};
\vertex (a11) at (-0.5*1.2,.75*1.2){};
\vertex (a12) at (-0.5*1.2,-0.75*1.2){};
\vertex (a13) at (0.5*1.2,.75*1.2){};
\vertex (a14) at (0.5*1.2,-0.75*1.2){};
\diagram{
(a1)--[ultra thick](a2),
(a3)--[ultra thick, hred](a4),
(a5)--[ultra thick, hred](a6),
(a7)--[ultra thick](a8),
(a13)--[scalar,ultra thick,hgreen](a14),
(a11)--[scalar,ultra thick,hgreen](a12),
};
\end{feynman}
\end{tikzpicture}
}
}

\newcommand{\diskSixpoint}[6]{ {
\begin{tikzpicture}[baseline=-2pt,squarednode/.style={rectangle, draw=black, fill=black!5, very thick, minimum size=5mm}]
\begin{feynman}
\vertex(mid1) at (0,0) {};
\draw (mid1) circle(0.51);
\vertex[dot,scale=0.9] (a4) at (-.5,0) {};
\vertex[dot,scale=0.9] (a3) at (-0.25,0.4330127) {};
\vertex[dot,scale=0.9] (a2) at (0.25,0.4330127) {};
\vertex[squarednode,scale=0.25] (a1) at (.5,0) {};
\vertex[dot,scale=0.9] (a6) at  (0.25,-0.4330127) {};
\vertex[dot,scale=0.9] (a5) at (-0.25,-0.4330127){};
\diagram{
(a#1)--[ultra thick](a#2),
(a#3)--[ultra thick, hred](a#4),
(a#5)--[ultra thick, hred](a#6),
};
\end{feynman}
\end{tikzpicture}
}
}

\newcommand{\diskSixpointAB}[6]{ {
\begin{tikzpicture}[baseline=-2pt,squarednode/.style={rectangle, draw=black, fill=black!5, very thick, minimum size=5mm}]
\begin{feynman}
\vertex(mid1) at (0,0) {};
\draw (mid1) circle(0.51);
\vertex[dot,scale=0.9] (a4) at (-.5,0) {};
\vertex[dot,scale=0.9] (a5) at (-0.25,0.4330127) {};
\vertex[dot,scale=0.9] (a6) at (0.25,0.4330127) {};
\vertex[squarednode,scale=0.25] (a1) at (.5,0) {};
\vertex[dot,scale=0.9] (a2) at  (0.25,-0.4330127) {};
\vertex[dot,scale=0.9] (a3) at (-0.25,-0.4330127){};
\diagram{
(a#1)--[ultra thick,quarter left,looseness=0.3](a#2),
(a#3)--[ultra thick,quarter right,looseness=0.3](a#4),
(a#5)--[ultra thick,quarter right,looseness=0.3](a#6),
};
\end{feynman}
\end{tikzpicture}
}
}

 \def\draftnote#1{{\color{red}\it #1}} 
\def\sect#1{section~\ref{#1}}

\def\spa#1.#2{\left\langle#1\,#2\right\rangle}
\def\spb#1.#2{\left[#1\,#2\right]}
\def\spash#1.#2{\spa{\smash{#1}}.{\smash{#2}}}
\def\spbsh#1.#2{\spb{\smash{#1}}.{\smash{#2}}}
\def\sand#1.#2.#3{%
\left\langle\smash{#1}{\vphantom1}^{-}\right|{#2}%
\left|\smash{#3}{\vphantom1}^{-}\right\rangle}
\def\sandpp#1.#2.#3{%
\left\langle\smash{#1}{\vphantom1}^{+}\right|{#2}%
\left|\smash{#3}{\vphantom1}^{+}\right\rangle}
\def\sandpm#1.#2.#3{%
\left\langle\smash{#1}{\vphantom1}^{+}\right|{#2}%
\left|\smash{#3}{\vphantom1}^{-}\right\rangle}
\def\sandmp#1.#2.#3{%
\left\langle\smash{#1}{\vphantom1}^{-}\right|{#2}%
\left|\smash{#3}{\vphantom1}^{+}\right\rangle}

\def\eqn#1{eq.~(\ref{#1})}

\def\be{\begin{equation}}
\def\ee{\end{equation}}
\def\bea{\begin{eqnarray}}
\def\eea{\end{eqnarray}}
\def\ba{\begin{eqnarray}}
\def\ea{\end{eqnarray}}

 \definecolor{MattOrange}{rgb}{1.0,0.4,0.2}

\begin{document}

\preprint{
}

\author{Nicolas H. Pavao}
\affiliation{Department of Physics and Astronomy, Northwestern
  University, Evanston, Illinois 60208, USA}

\title{Effective observables for electromagnetic duality
\\
from novel amplitude decomposition}

\begin{abstract}
We introduce a decomposition of $D$-dimensional vector amplitudes in terms of building blocks that preserve all partial amplitude relations of the parent gauge-theory. Using this decomposition, we derive a new set of amplitude relations between nonlinear sigma model pions and the pure-scalar sector of Yang-Mills-scalar theory. These new relations indicate an equivalence between Born-Infeld duality invariance and that of Einstein-Maxwell theory, considering their known double-copy construction from the two scalar theories with Yang Mills. This observation motivates a general framework for constructing duality-invariant effective field theory observables at any multiplicity using the building blocks in our amplitude decomposition. Finally, we conjecture a subtle relationship between these duality-invariant building blocks and the bosonic sector of $\mathcal{N}=4$ super Yang-Mills. \end{abstract}

\maketitle

\section{Introduction}\label{sec:intro}
Electromagnetic duality, sometimes called the $U(1)$ duality, is a continuous symmetry of Abelian vector fields that leaves the equations of motion invariant. It is a classically conserved symmetry of the Standard Model, albeit anomalous \cite{Schwinger:1951nm}. To be precise, a theory is said to exhibit electromagnetic duality if the equations of motion are symmetric under the following $U(1)$ rotation,
\begin{equation}\label{eq:phase}
F^{\mu\nu}+iG^{\mu\nu} \rightarrow e^{i\alpha } (F^{\mu\nu}+iG^{\mu\nu})\,,
\end{equation}
where $G^{\mu\nu}$ is defined as,
\begin{equation}
G^{\mu\nu} \equiv \epsilon^{\mu\nu\rho\sigma}\frac{\partial \mathcal{L}}{\partial F^{\rho\sigma}}\,,
\end{equation}
and $F^{\mu\nu}$ is the standard Abelian field strength of electromagnetism. Theories that respect this symmetry, along with their observable quantities, are called \textit{duality-invariant}. In recent years, the $U(1)$ duality has seen a resurgence in the literature for being linked to good ultraviolet behavior in point-like theories of quantum gravity \cite{Bossard:2012xs,Carrasco:2013ypa,Bern:2013uka}. Due to this renewed interest, there has been tremendous progress in understanding electromagnetic duality in the framework of on-shell methods \cite{Cachazo:2014xea,Bern:2017rjw,Novotny:2018iph,Elvang:2020kuj}. In this paper, we will show that many duality-invariant scattering amplitudes, including those of Born-Infeld (BI) theory \cite{Born:1934gh,Schrodinger:1935oqa}, actually belong to a unified class of effective field theory (EFT) observables that can be constructed from a small set of gauge theory building blocks.

The simplest duality-invariant theory is Maxwell's theory; of course, due to the linearity of the equations of motion, the S-matrix for this theory is trivial. However, one can construct an interacting Maxwellian theory, while preserving duality invariance, by minimally coupling the photon to dynamic spacetime \cite{Gibbons:1995ap}, and introducing any number of $U(1)$ gauge fields, 
\begin{equation}\label{eq:EMf}
\mathcal{L}_{\text{EMf}} = \sqrt{-g}\, \Big(R-\frac{1}{4} \sum_{I} F_{\mu\nu}^I F^{I\mu\nu}\Big)\,.
\end{equation}
Since the photon stress tensor is itself invariant under duality rotations, the Lagrangian above also exhibits electromagnetic duality. The additional $U(1)$ gauge fields only interact with each other through minimal coupling to the background; thus, they are in some sense flavored. For this reason, the model above has been referred to as flavored Einstein-Maxwell (EMf) theory \cite{Zhou:2019gtk}. 

The scattering amplitudes generated by EMf theory have a couple of notable properties. First, they can be constructed by double-copying Yang-Mills (YM) with pure-scalar amplitudes in Yang-Mills-scalar (YMS) theory. For recent reviews on double-copy construction see \cite{Bern:2019prr, Adamo:2022dcm, Bern:2022wqg}. Second, the on-shell phase symmetry in \eqn{eq:phase}, is associated with a conserved chiral charge \cite{Novotny:2018iph}. This guarantees helicity conservation at the level of amplitudes. With all momentum taken to be outgoing, this forces the following constraint on the tree-level S-matrix elements,
\begin{equation}\label{eq:EMfDC}
\mathcal{M}_{(n_+,n_-)}^{\text{EMf}}=\mathcal{A}^{\text{YM}}_{(n_+,n_-)}\otimes \mathcal{A}^{\text{YMS}} = 0, \qquad n_+\neq n_-\,,
\end{equation}
where $n_+$ and $n_-$ indicate the number of positive and negative on-shell helicity states, respectively, and $\otimes$ is the field theory Kawai-Lewellen-Tye (KLT) momentum kernel \cite{Kawai:1985xq,Bern:1998sv,Bjerrum-Bohr:2010pnr} that carries out double-copy construction \cite{Bern:2008qj, Bern:2010ue} at the level of partial amplitudes. Note that this constraint on the S-matrix of EMf theory is true even for nonvanishing $\text{N}^k\text{MHV}$ configurations of YM -- this is an example of KLT orthogonality \cite{BjerrumBohr2010ta, BjerrumBohr2010yc}. 

While the amplitudes generated by \eqn{eq:EMfDC} actually correspond to photons coupled to $\mathcal{N}=0$ axion-dilaton-supergravity, the helicity selection rule above still holds. This has been shown both with on-shell methods \cite{Cachazo:2014xea}, and at the level of the action \cite{Gibbons:1995ap,Babaei-Aghbolagh:2013hia}. Since the selection rule applies for all flavor configurations of the YMS amplitudes, any linear combination of $\mathcal{A}^{\text{YMS}}$ independent of color ordering must also produce duality-invariant observables when fed into the right-hand side of \eqn{eq:EMfDC}.

An example of one such linear decomposition that we will derive, is the relationship between nonlinear sigma model (NLSM) pion amplitudes and flavor permutations of YMS amplitudes:
\begin{equation}\label{eq:firstNLSMtoYMS}
A^{\text{NLSM}}_{2k} = (-1)^{k-1}\sum_{\rho\in S^{2|(k-1)}_{(ij)^c}}s_{(\rho)} A^{\text{YMS}}_{(ij)(\rho)}\,.
\end{equation}
The formalism in \eqn{eq:firstNLSMtoYMS} will be described in \sect{sec:RABD}. Once we have identified the class of linear relations to which the above scalar identity belongs, then we will use the elements of the expansion to construct higher derivative (HD) observables that maintain duality invariance at any desired multiplicity. 
\section{Reducible Amplitude Block Decomposition}\label{sec:RABD}
First we introduce our sector decomposition for gauge theory amplitudes. In this paper we focus on the decomposition applied to $D$-dimensional Yang-Mills amplitudes for the purpose of constructing duality-invariant matrix elements. However, the procedure that follows applies to any HD generalization. 

By multiplying and dividing contact terms by propagators, local gauge theory amplitudes at $n$-point can be expressed as a sum over trivalent graphs, $\Gamma^{(3)}_n$, each one weighted by color factors, $C_g$, and kinematic weights, $N_g$,
\begin{equation}
\mathcal{A}^{\text{vec}}_n = \sum_{g\in \Gamma^{(3)}_n} \frac{C_gN_g}{D_g}\,,
\end{equation}
where $D_g$ encodes the propagator structure of a given graph, $g$. While manifestly local, this representation obscures gauge invariance of the full color-dressed amplitude, $\mathcal{A}_n^{\text{vec}}$. To make gauge invariance manifest, the color factors can be decomposed into a sum over color traces, which gives rise to the so called trace-basis representation \cite{Dixon:1996wi}, 
\begin{equation}
\mathcal{A}^{\text{vec}}_n = \sum_{\sigma\in S^{n}/Z_n} \text{Tr}(T^{\sigma_1}T^{\sigma_2}\cdots T^{\sigma_n})A^\text{vec}_{(\sigma)}\,,
\end{equation}
where $T^{\sigma_i}$ is the group generator assigned to the $i^{th}$ particle. Since the group theory trace structures are linearly-independent objects in color space, they are weighted by kinematic functions, $A^{\text{vec}}_{(\sigma)}$, that are themselves gauge invariant. For this reason, these kinematics functions are referred to as \textit{partial amplitudes}. 

In general, the partial amplitudes are functions of $D$-dimensional dot products, $(k_ak_b), (k_a\epsilon_b)$, and $(\epsilon_a\epsilon_b)$. In this paper, we will use the shorthand notation for each of these dot products,
\begin{equation}
\begin{split}
s_{(a_1 b_1)...(a_n b_n)} &= (k_{a_1}k_{b_1}) \dots (k_{a_n}k_{b_n})\,,
\\
\epsilon_{(a_1 b_1)...(a_n b_n)} &= (\epsilon_{a_1}\epsilon_{b_1}) \dots (\epsilon_{a_n}\epsilon_{b_n})\,,
\\
\kappa^{(b)}_{a_1...a_n} &= (k_{a_1}+\dots +k_{a_n})\epsilon_b\,.
\end{split}
\end{equation}
Due to on-shell constraints and polarization transversality, $\kappa^{(a)}_{a}=0$, dot products that include factors of momenta depend on the choice of kinematic basis used to represent the amplitude. In contrast, the polarization products, $\epsilon_{(ab)}$, are basis independent. This allows us to express any amplitude, $A^{\text{vec}}_{(\sigma)}$, as a sum over distinct polarization products,
\begin{equation}\label{eq:pureVecRABD}
A_{(\sigma)}^\text{vec} = \sum_{k=0}^{\lfloor |\sigma|/2\rfloor}\sum_{\rho \in S^{2|k}_{\sigma}}\epsilon_{(\rho)} \Delta_{(\sigma)}^{(\rho)}\,.
\end{equation}
We refer to this construction as the \textit{Reducible-Amplitude Block-Decomposition} (RABD). The sum is over the elements in $S^{2|k}_{\sigma}$, which contains all permutations of $k$ pairs appearing in the ordered set of external legs, $\sigma$. For example, the elements appearing in $S^{2|1}_{(1234)}$ and $S^{2|2}_{(1234)}$ would be,
\begin{align}
S^{2|1}_{(1234)} &= \Big\{(12),(13),(14),(23),(24),(34)\Big\}\,,
\\
S^{2|2}_{(1234)} &= \Big\{(12)(34),(13)(24),(14)(23)\Big\}\,.
\end{align}
Now that we have stripped away all the basis independent parts of the amplitude, the kinematic functions $ \Delta_{(\sigma)}^{(\rho)}$ are defined implicitly as the coefficients of polarization products, $\epsilon_{(\rho)}$. Thus, they are functions exclusively of $s_{ab}$ and $\kappa^{(b)}_{a_1...a_n}$ kinematics,
\begin{equation}
\Delta_{(\sigma)}^{(\rho)}\equiv \Delta_{(\sigma)}^{(\rho)}\left(\{s_{ab},\kappa^{(b)}_{a}\}\right)\,.
\end{equation} 
This is in a similar spirit to the decomposition of \cite{Dong:2021qai}, which expanded amplitudes in terms of lower spin kinematics, with each term weighted by a gauge-invariant prefactor. In contrast, in \eqn{eq:pureVecRABD} we extract polarization dot-products in such a way that the RABD unequivocally violates gauge invariance term-by-term. In return for temporarily surrendering gauge invariance, the building blocks, $ \Delta_{(\sigma)}^{(\rho)}$, inherit all partial amplitude relations \cite{Bern:2008qj,Kleiss:1988ne,DelDuca:1999rs} of the parent amplitude. Take for example a partial amplitude $A^{\text{vec}}_{(\sigma)}$ that satisfies the fundamental Bern-Carrasco-Johansson (BCJ) relations \cite{Bern:2008qj,Feng:2010my} on the external legs:
\begin{equation}\label{eq:fieldTheoryPreserving}
\begin{split}
&\sum_{i=2}^{n-1} s_{2...i|1}A^{\text{vec}}_{(2,...,i,1,i+1,...,n)}=0
\\
\Rightarrow& \sum_{i=2}^{n-1} s_{2...i|1}\left(\sum_{k=0}^{\lfloor |\sigma|/2\rfloor}\sum_{\rho \in S^{2|k}_{\sigma}}\epsilon_{(\rho)}\Delta^{(\rho)}_{(\sigma)}\right)=0
\\
\Rightarrow& \sum_{k=0}^{\lfloor |\sigma|/2\rfloor}\sum_{\rho \in S^{2|k}_{\sigma}}\epsilon_{(\rho)} \left(\sum_{i=2}^{n-1} s_{2...i|1}\Delta^{(\rho)}_{(\sigma)}\right)=0
\\
\Rightarrow& \sum_{i=2}^{n-1} s_{2...i|1}\Delta^{(\rho)}_{(2,...,i,1,i+1,...,n)} =0\,,
\end{split}
\end{equation}
where $s_{2...i|1} =  (k_2+\cdots +k_i) k_1$, and to simplify notation we have taken $\sigma=(2,...,i,1,i+1,...,n)$ at intermediate steps. The takeaway is that since the distinct polarization sectors, $\epsilon_{(\rho)}$, are independent of the leg ordering, $\sigma$, all field theory relations that act on $\sigma$ are inherited by the building blocks. This property is the key that will permit us to construct generalized building blocks for duality invariance using the KLT momentum kernel.  

To gain familiarity with the RABD in expanded form, below we provide an explicit example of \eqn{eq:pureVecRABD}, where a four-point partial amplitude with leg ordering $\sigma=(1234)$ can be expressed in terms of our building blocks as follows,
\begin{equation}
\begin{split}
A_{(\sigma)}^\text{vec} &=\epsilon_{(12)}\Delta^{(12)}_{(\sigma)} + \epsilon_{(34)}\Delta^{(34)}_{(\sigma)} +\epsilon_{(12)(34)}\Delta^{(12)(34)}_{(\sigma)} 
\\
&+ \epsilon_{(13)}\Delta^{(13)}_{(\sigma)} + \epsilon_{(24)}\Delta^{(24)}_{(\sigma)}+ \epsilon_{(13)(24)}\Delta^{(13)(24)}_{(\sigma)} 
 \\
& +\epsilon_{(14)}\Delta^{(14)}_{(\sigma)}+\epsilon_{(23)}\Delta^{(23)}_{(\sigma)} + \epsilon_{(14)(23)}\Delta^{(14)(23)}_{(\sigma)} 
\\
& + \Delta^{(\varnothing)}_{(\sigma)}  \,.
\end{split}
\end{equation} 
Considering the proliferation of notation, we will utilize a diagrammatic form of these building blocks. To construct the diagrams for $\Delta_{(\sigma)}^{(\rho)}$, first draw a set of labeled points on a circle according to the ordered list, $\sigma$. To avoid numbering all the points, draw a box to denote the first entry in $\sigma$, and orient the diagrams in a counterclockwise direction. Then draw a line for every pair appearing in the set $\rho$. Below are some eight-point examples:
\begin{equation}\label{eq:exDiagrams}
\begin{split}
\Delta_{(12345678)}^{(15)(38)} &= \diskEightpointAB{1}{5}{3}{8}{}{}{}{}\,,
\\
\Delta_{(12345678)}^{(14)(36)(58)} &= \diskEightpointAB{1}{4}{3}{6}{5}{8}{}{}\,,
\\
\Delta_{(12345678)}^{(12)(47)(58)(36)} &=\diskEightpointABcurve{4}{7}{1}{2}{5}{8}{6}{3}\,.
\end{split}
\end{equation}
When the kinematic functions are stripped from Yang-Mills amplitudes and $|\sigma| = |\rho|$, these diagrams are precisely the pure-scalar YMS amplitudes generated by the dimensional-reduction of Yang-Mills, such that $\Delta_{(\sigma)}^{(\rho)}=A^{\text{YMS}}_{(\sigma|\rho)}$. Under these conditions the diagrams map onto those used by Cachazo, He and Yuan (CHY) in \cite{Cachazo:2014xea}. 

In the proceeding sections, it will become useful to weight a given pair, $(ab)\subset \rho$, appearing in a $\Delta^{(\rho)}_{(\sigma)}$ diagram by dot-products, $s_{ab}$ and $\kappa^{(b)}_a$. This is encoded in our diagrams by coloring the $(ab)$ weighted line and $a^{th}$ momentum dot, respectively. Take for example the second diagram in \eqn{eq:exDiagrams}, multiplied by a $\kappa^{(6)}_3\kappa^{(8)}_5 $ kinematic factor,
\begin{equation}
\begin{split}
\kappa^{(6)}_3\kappa^{(8)}_5 \Delta_{(12345678)}^{(14)(36)(58)} =\diskEightpointWithDot{1}{4}{}{}{3}{6}{5}{8}\,,
\end{split}
\end{equation}
or similarly weighted by a $s_{36}s_{58}$ product of Mandelstams,
\begin{equation}
\begin{split}
s_{36}s_{58} \Delta_{(12345678)}^{(14)(36)(58)} =\diskEightpoint{1}{4}{}{}{3}{6}{5}{8}\,.
\end{split}
\end{equation}
For any amplitude that can be expressed in terms of the RABD of \eqn{eq:pureVecRABD}, we can also use $\Delta^{(\rho)}_{(\sigma)}$ to construct mixed scalar-vector amplitudes that are themselves gauge invariant. This is done by taking pairs of polarization vectors that appear in the partial amplitude, $A^{\text{vec}}_{(\sigma)}$, and giving them nonvanishing values orthogonal to the momentum subspace such that
\begin{equation}
\epsilon_{(a_ib_j)} = \delta_{ij}, \qquad a_i,b_j \in \sigma\,.
\end{equation}
This is equivalent to acting a product of transmutation operators \cite{Cheung:2017ems} on the parent vector amplitude:
\begin{equation}
A^{\varphi+\text{vec}}_{(\sigma|\beta)} = \prod_{i=1}^k \mathcal{\partial}_{\epsilon_{(a_ib_i)}}A^{\text{vec}}_{(\sigma)}\,,
\end{equation}
where $\beta=\{(a_1b_1)\dots (a_kb_k)\}$ is a list of $k$ scalar pairs. From the RABD, which factors out all the $\epsilon_{(ab)}$ terms, we can see immediately which terms survive:
\begin{equation}\label{eq:mixedRABD}
A_{(\sigma|\beta)}^{\varphi+\text{vec}} = \sum_{k=0}^{\lfloor |\beta|^c/2 \rfloor} \sum_{\rho \in S^{2|k}_{\beta^c}} \epsilon_{(\rho)} \Delta_{(\sigma)}^{(\beta \cup \rho)}\,,
\end{equation}
where $\beta^c \equiv \sigma \setminus \beta$ is the complement set of vectors that remain. Take for example the following two-scalar four-vector amplitude expressed in terms of our diagrams:
\begin{equation} \label{eq:sixPointEx}
\begin{split}
&A_{(123456|(12))}^{\varphi+\text{vec}}=\diskSixpointStraight{1}{2}{}{}{}{}
\\
&+\epsilon_{(34)}\diskSixpointStraight{1}{2}{3}{4}{}{}+\epsilon_{(56)}\diskSixpointStraight{1}{2}{5}{6}{}{}+\epsilon_{(34)(56)}\diskSixpointStraight{1}{2}{5}{6}{3}{4}
\\
&+\epsilon_{(35)}\diskSixpointStraight{1}{2}{3}{5}{}{}+\epsilon_{(46)}\diskSixpointStraight{1}{2}{4}{6}{}{}+\epsilon_{(35)(46)}\diskSixpointStraight{1}{2}{4}{6}{3}{5}
\\
&+\epsilon_{(36)}\diskSixpointStraight{1}{2}{3}{6}{}{}+\epsilon_{(45)}\diskSixpointStraight{1}{2}{4}{5}{}{}+\epsilon_{(36)(45)}\diskSixpointStraight{1}{2}{4}{5}{3}{6}\,.
\end{split}
\end{equation}In the event that $\beta = \varnothing$, then this generalized block decomposition in \eqn{eq:mixedRABD} reproduces the pure vector form of \eqn{eq:pureVecRABD}. It is important to note that the building blocks in the RABD are not necessarily themselves gauge-invariant. To maintain gauge invariance in the partial amplitude, these kinematic functions, $\Delta_{(\sigma)}^{(\rho)}$, must be related in the following way,
 \begin{equation}\label{eq:GIrelA}
\Delta_{(\sigma)}^{(\rho)}\Big|_{\epsilon_i\rightarrow k_i} =-
\sum_{j \in \rho^c} \kappa^{(j)}_i \Delta_{(\sigma )}^{(\rho\cup (ij))} \,.
\end{equation}
This Ward identity follows from the property that individual products of $\epsilon_{(\rho)}$ label distinct kinematic sectors of the amplitude. Since taking a particular $\epsilon_i \rightarrow k_i$ will have the following effect,
\begin{equation}
\epsilon_{(ab)}\Big|_{\epsilon_i \rightarrow k_i} = \begin{cases} 
\epsilon_{(ab)} & i\neq a,b
\\
\kappa^{(a)}_b & i =b
\\
\kappa^{(b)}_a & i =a
\end{cases},
\end{equation}
individual sectors labeled by the surviving $\epsilon_{(\rho)}$ must then independently vanish. This is captured by the Ward identity of \eqn{eq:GIrelA}. The following is a diagrammatic example of this relationship when taking $\epsilon_3\rightarrow k_3$ in the first diagram of \eqn{eq:sixPointEx}:
\begin{equation}
\diskSixpointStraight{1}{2}{}{}{}{}\,\Bigg|_{\epsilon_3\rightarrow k_3} =-\left(\diskSixpointPol{1}{2}{3}{4}{}{}+\diskSixpointPol{1}{2}{3}{5}{}{}+\diskSixpointPol{1}{2}{3}{6}{}{}\right)\,.
\end{equation}
Furthermore, since $\kappa^{(b)}_{a}|_{\epsilon \rightarrow k} = s_{ab}$, we can apply the Ward identity recursively, and obtain the following expansion of even-point building blocks when all polarizations are projected along their momenta:
\begin{equation}\label{eq:GIrelB}
\Delta_{(\sigma)}^{(\rho)}\Big|_{\epsilon\rightarrow k} =
(-1)^{|\rho^c /2|}\sum_{\tau \in S_{\rho^c}^2} s_{(\tau)}\Delta_{(\sigma)}^{(\rho \cup \tau )}\,.
\end{equation}
A diagrammatic description of this additional identity acting on the first term of \eqn{eq:sixPointEx} is provided below,
\begin{equation}
\diskSixpointStraight{1}{2}{}{}{}{}\,\Bigg|_{\epsilon\rightarrow k} =\diskSixpoint{1}{2}{3}{4}{5}{6}+\diskSixpoint{1}{2}{3}{5}{4}{6}+\diskSixpoint{1}{2}{3}{6}{5}{4}\,.
\end{equation}
While there are many applications of this decomposition, some of which we will discuss in \sect{sec:Discussion}, for the remainder of this paper we will focus on the special case of Yang-Mills building blocks. This particular set of kinematics will allow us to gain insight about structure that underlies NLSM pion amplitudes and make incredibly robust statements about the construction of duality-invariant observables at all multiplicity. 
\section{New Amplitude Relations}\label{sec:AmpRel}
Now that we have described the construction of the RABD, and identified a handful of relationships between the building blocks required by gauge invariance, we are prepared to express NLSM amplitudes in terms of YMS amplitudes. The first step is to note that there exists a mapping between gluons in $(2d+1)$-dimensions and pions in $d$-dimensions \cite{Cheung:2017yef}. This mapping can be implemented with a variety of equivalent operations, but for our purposes we will employ a standard dimensional-reduction procedure at the level of kinematic variables. By considering the manifestly color-dual NLSM action in \cite{Cheung:2016prv}, it was shown in \cite{Cheung:2017yef} that one can select any two legs, $(ij)$, and plug in the following polarizations:
\begin{equation}\label{eq:genDim}
\mathcal{E}_a = \begin{cases} (\vec{0},1,\vec{0}) & a = i,j
\\
(k^\mu_a,0,ik^\mu_a) & a \neq i,j
\end{cases}\,,
\end{equation}
where the first and last entries are $d$-dimensional vectors, and the $(2d+1)$-dimensional momenta are restricted to the $d$-dimensional subspace, $\mathcal{K}_a = (k^\mu_a,0,\vec{0})$. We will use a generalization of this dimensional-reduction procedure in \sect{sec:DIObs} to target a wide class duality-invariant matrix elements. 

By plugging polarizations of \eqn{eq:genDim} into our $D$-dimensional expansion of \eqn{eq:pureVecRABD}, we can see that this choice of polarizations targets a very specific block in the RABD. From which we conclude,
\begin{equation}\label{eq:NLSMfromblock}
A^{\text{NLSM}}_{(\sigma)} = \Delta^{(ij)}_{(\sigma)} \Big|_{\epsilon\rightarrow k} \qquad \forall (ij) \subset \sigma\,,
\end{equation}
where $\Delta^{(ij)}_{(\sigma)}$ is a single building block in the RABD of Yang-Mills. 

Before proceeding, we note that these color-dual building blocks in \eqn{eq:NLSMfromblock} are special in that they lack contributions from contact diagrams, and thus serve as a prime candidates for constructing kinematic structure constants. The lack of contacts is due to a simple mass-dimension and little-group scaling argument. At $n$-point, YM numerators carry mass-dimension $\big[N^{\text{YM}} \big]= n-2$; likewise, the $\Delta^{(ij)}_{(\sigma)}$ building block contains $n-2$ polarizations, by definition. Thus, the kinematic numerators for this building block, $N^{\text{YM}}_{(ij)}$, are exclusively functions of $\kappa^{(b)}_a$ kinematics:
\begin{equation}
N^{\text{YM}}_{(ij)}\equiv N^{\text{YM}}_{(ij)}\left(\{\kappa^{(b)}_a\}\right)
\end{equation}
Since the numerators for $\Delta^{(ij)}_{(\sigma)}$ contain no factors of $s_{ab}$ to cancel poles in the denominators, the entire building block is necessarily cubic\footnote{This is not dissimilar from the argument presented in \cite{Elvang:2013cua} for why MHV Yang-Mills amplitudes are fundamentally cubic at tree-level.}. Furthermore, since the building blocks themselves satisfy BCJ relations, the numerators should permit a color-dual representation \cite{Bern:2008qj}. Indeed, we find the following color-dual dressing for the half-ladder master diagram, $N^{\text{YM}}_{(i|\tau|j)}$, 
\begin{equation}\label{eq:HLblock}
N^{\text{YM}}_{(i|\tau|j)}\equiv N^{\text{YM}}_{(ij)} \Bigg(\! \halfLadder{} \! \Bigg)= \prod_{k=2}^{n-1} \kappa^{(\tau_k)}_{i\tau_2...\tau_{k-1}}\,.
\end{equation}
If we demand the numerators be polynomial in kinematics, this dressing is unique. Each term in the product expansion of \eqn{eq:HLblock} serves as a cubic structure constant, like those found for the MHV sector in \cite{Monteiro2011pc}. From this dressing, the building block needed in \eqn{eq:NLSMfromblock} can be recovered by contracting the $n-2$ of vector indices with the doubly ordered propagator matrix \cite{Vaman:2010ez}, $m_{(\sigma|\tau)}$, of bicolored scalar amplitudes \cite{Bjerrum-Bohr:2012kaa},
\begin{equation}
 \Delta^{(ij)}_{(\sigma)} = \sum_{\tau \in S^{n-2}} m_{(\sigma|\tau)} N^{\text{YM}}_{(i|\tau|j)}\,.
\end{equation}
As a consistency check, since $m_{(\sigma|\tau)}$ has no polarization dependence, taking $\epsilon \rightarrow k$ should yield a valid expression for half-ladder numerators of NLSM at $n$-point. Indeed, we find this replacement gives rise to precisely the dressings found in Ref.~\cite{Carrasco:2016ldy}.

Returning to our originally stated goal, the final step is to express $\Delta^{(ij)}_{(\sigma)}$ in terms of the pure-scalar building blocks when taking all $\epsilon \rightarrow k$. The procedure for this was shown in \eqn{eq:GIrelB}. With this step, we find the promised linear decomposition that maps YMS amplitudes to NLSM pions:
\begin{equation}\label{eq:YMStoNLSM}
A^{\text{NLSM}}_{2k} = (-1)^{k-1}\sum_{\rho\in S^{2|(k-1)}_{(ij)^c}} s_{(\rho)} A^{\text{YMS}}_{(ij)(\rho)}\,.
\end{equation}
Since \eqn{eq:YMStoNLSM} holds for any choice of $(ij)$, below is just one of many valid expressions for the eight-point NLSM partial amplitude:
\begin{equation}\label{eq:eightPointRel}
\begin{split}
|A_8^{\text{NLSM}}|&=\diskEightpoint{1}{5}{2}{3}{4}{6}{7}{8}+\diskEightpoint{1}{5}{2}{3}{4}{8}{7}{6}+\diskEightpoint{1}{5}{2}{8}{4}{3}{7}{6}+\diskEightpoint{1}{5}{2}{6}{4}{3}{7}{8}
\\
&+\diskEightpoint{1}{5}{2}{4}{3}{8}{7}{6}+\diskEightpoint{1}{5}{2}{4}{3}{6}{7}{8}+\diskEightpoint{1}{5}{2}{3}{4}{7}{6}{8}+\diskEightpoint{1}{5}{2}{7}{4}{3}{6}{8}
\\
&+\diskEightpoint{1}{5}{2}{8}{4}{6}{7}{3}+\diskEightpoint{1}{5}{2}{4}{3}{7}{6}{8}+\diskEightpoint{1}{5}{2}{8}{4}{7}{6}{3}+\diskEightpoint{1}{5}{2}{7}{4}{6}{3}{8}
\\
&+\diskEightpoint{1}{5}{2}{6}{4}{7}{3}{8}+\diskEightpoint{1}{5}{2}{6}{4}{8}{7}{3}+\diskEightpoint{1}{5}{2}{7}{4}{8}{3}{6}
\\
&=\sum_{\rho\in S^{2|3}_{(15)^c}} s_{(\rho)} A^{\text{YMS}}_{(15)(\rho)}\,.
\end{split}
\end{equation}
Taking a closer look, we note that the last five terms actually vanish! This was pointed out in \cite{Cachazo:2014xea}. The reason for this disappearance is due to the available vertices in YMS theory. Consider the Lagrangian for a YMS theory with $M$-complex scalars coupled to a $d$-dimensional gluon,
\begin{equation}\label{eq:ymsLag}
\mathcal{L} = -\frac{1}{4}F_{\mu\nu}F^{\mu\nu} +\frac{1}{2}( D_\mu \varphi^i) (D^\mu \bar{\varphi}^i) - \frac{1}{4}[\varphi^i, \bar{\varphi}^j ][\bar{\varphi}^{i}, \varphi^{j}]\,,
\end{equation}
where there is an implied trace over color indices. Given the four-point contact, only two flavor lines are permitted to cross at a point; and at tree level, all sub-areas bounded by the flavor lines must touch the exterior. 

To our understanding, it is an open question as to how many nonvanishing graphs can be drawn at multiplicity $2k$, given these constraints. We call this combinatorial problem the \textit{pizza crust problem}, as it amounts to the practical dilemma of determining how many different ways one can arrange $k$ ordered slicings of a pizza pie, such that every connected area has at least one edge of crust. Pies with crustless pieces must vanish when mapped to YMS amplitudes. Below are some slicings that lead to \textit{vanishing} YMS amplitudes,
\begin{equation}
\diskEightpointAB{1}{4}{2}{7}{3}{6}{5}{8}\qquad \diskSixpointAB{1}{4}{2}{5}{6}{3}\qquad \diskEightpointABcurve{4}{7}{1}{2}{5}{8}{6}{3}\,,
\end{equation}
whereas the following are some examples of crusted slicings, which correspond to \textit{nonvanishing} elements in the space of distinct YMS amplitudes:
\begin{equation}
\diskEightpointAB{1}{4}{2}{7}{5}{3}{6}{8}\qquad \diskEightpointABcurve{1}{2}{3}{4}{5}{7}{6}{8}\qquad \diskEightpointAB{1}{5}{2}{8}{3}{7}{6}{4}\,.
\end{equation}
At high multiplicity, nonvanishing YMS amplitudes will be in short supply due to the limited availability of flavor-mixing contacts. Consider the demonstrative example of a building block that appears in the expansion of a 16-point YM amplitude (shown below), for which half the polarizations are stripped into the $\epsilon_{(\rho)}$ prefactor. The Ward identity of \eqn{eq:GIrelA} suggests we will need to sum over all $(8-1)!! = 105$ valid permutations when taking $\epsilon\rightarrow k$. However, remarkably, one finds that all but four of the possible YMS flavor configurations vanish:
\begin{equation}\label{eq:sixteenPointRel}
\diskSixteenpointB{}{}{}{}{}{}{}{} \Bigg|_{\epsilon\rightarrow k}=\,\diskSixteenpointB{2}{1}{8}{7}{11}{5}{4}{14}\,+\,\diskSixteenpointC{2}{1}{7}{5}{11}{8}{4}{14}\,+\,\diskSixteenpointC{1}{14}{4}{2}{8}{7}{11}{5}\,+\,\diskSixteenpointA{1}{14}{4}{2}{7}{5}{11}{8}\,.
\end{equation}
This striking example suggests that the space of nonvanishing YMS amplitudes grows significantly more slowly than the dimension of $S^{2|(|\sigma|/2)}_{\sigma}$, 
\begin{equation}
||S^{2|(|\sigma|/2)}_{\sigma}|| = (|\sigma|-1)!!\,, 
\end{equation}
which serves as an upper bound. As we will argue in \sect{sec:DIObs}, determining the exact dimension of the space of nonvanishing YMS with $k$ flavors at $2k$-point could have interesting implications for expressing NLSM amplitudes in terms of pure-scalar supersymmetric (SUSY) amplitudes. We will leave the solution to the pizza crust problem as a potential direction of future study. 

Putting this all together, the pion decomposition in \eqn{eq:eightPointRel} can be reduced to a sum over just ten nonvanishing diagrams:
\begin{equation}
\begin{split}
|A_8^{\text{NLSM}}|&=\diskEightpoint{1}{5}{2}{3}{4}{6}{7}{8}+\diskEightpoint{1}{5}{2}{3}{4}{8}{7}{6}+\diskEightpoint{1}{5}{2}{8}{4}{3}{7}{6}+\diskEightpoint{1}{5}{2}{6}{4}{3}{7}{8}
\\
&+\diskEightpoint{1}{5}{2}{4}{3}{8}{7}{6}+\diskEightpoint{1}{5}{2}{4}{3}{6}{7}{8}+\diskEightpoint{1}{5}{2}{3}{4}{7}{6}{8}+\diskEightpoint{1}{5}{2}{7}{4}{3}{6}{8}
\\
&+\diskEightpoint{1}{5}{2}{8}{4}{6}{7}{3}+\diskEightpoint{1}{5}{2}{4}{3}{7}{6}{8}\,.
\end{split}
\end{equation}
As a result of this transformation that maps YMS into NLSM pion amplitudes, we can see that BI is essentially a HD projection from the space of duality-invariant EMf amplitudes:
\begin{align}
\mathcal{M}^{\text{BI}} &= \mathcal{A}^{\text{YM}} \otimes \mathcal{A}^{\text{NLSM}}
\\
& \equiv A^{\text{YM}}_{(\sigma)} S_{[\sigma|\tau]} A^{\text{NLSM}}_{(\tau)}
\\
&\sim \sum_{\rho}s_{(\rho)}A^{\text{YM}}_{(\sigma)} S_{[\sigma|\tau]} A^{\text{YMS}}_{(\tau|\rho)}
\\
&= \sum_{\rho}s_{(\rho)} \mathcal{M}^{\text{EMf}}_{(\rho)}\,,
\end{align}
where $S_{[\sigma|\tau]}$ is the KLT momentum kernel \cite{Bern:1998sv,Bjerrum-Bohr:2010pnr}. In the third line, we have factored out the kinematic flavor weight, $s_{(\rho)}$, as it is independent of the color ordering. Reinstating the factors of $(-1)$, we conclude the following relationship between the amplitudes of the two theories:
\begin{equation}\label{eq:BIfromEMf}
\mathcal{M}^{\text{BI}}_{2k} =(-1)^{k-1}\sum_{\rho\in S^{2|k-1}_{(ij)^c}}s_{(\rho)} \mathcal{M}^{\text{EMf}}_{(ij)(\rho)}\,.
\end{equation}
This observation echoes a known insight from the CHY description of these two amplitudes. In the CHY paradigm, one expresses $n$-point amplitudes, $\mathcal{M}_n$, as integrals over Riemann spheres punctured at points, $\sigma_a$,  \cite{Cachazo:2013gna,Cachazo:2013iea},
\begin{equation}
\mathcal{M}_n = \int d \mu_n \,\mathcal{I}_n(k,\epsilon,\sigma)\,. 
\end{equation}
The integrands $\mathcal{I}_n$ are theory-dependent functions of kinematics, and the measure $d \mu_n$ gives the integral support on the $(n-3)!$ different configurations of $\sigma_a$ that solve the scattering equations \cite{Cachazo:2013gna}, 
\begin{equation}\label{eq:scatteringEquations}
\sum_{a\neq b}\frac{s_{ab}}{\sigma_a-\sigma_b} = 0\,,
\end{equation}
for $a,b\in\{1,2,\dots,n\}$. The CHY integrands for BI and EMf theory both contain a factor of $\text{Pf}{\,'}A $ \cite{Cachazo:2014xea}, which is a special function of $D$-dimensional kinematics. In four-dimensions, solutions to \eqn{eq:scatteringEquations} split into distinct sectors that can be labeled by different helicity configurations, $(n_+,n_-)$. However, the kinematic factor $\text{Pf}{\,'}A$ vanishes outside the $n_+ = n_- = n/2$ solution space \cite{Cachazo:2014xea}. In other words, both BI and EMf must conserve helicity, $n_+ = n_-$. Not surprisingly, $\text{Pf}{\,'}A$ also appears in the construction of both YMS and NLSM integrands \cite{Cachazo:2014xea,He:2016vfi}.

The insight of CHY is that both BI and EMf owe their duality invariance to the appearance of at least one factor of $\text{Pf}{\,'}A$ in their integrands. While this observation is certainly suggestive of \eqn{eq:YMStoNLSM} and \eqn{eq:BIfromEMf}, what we have shown is an apparently stronger relationship that relates amplitudes to amplitudes, post-integration, rather than identifying shared kinematic factors, pre-integration. Equipped with an amplitude level description of the correspondence between BI and EMf, we are prepared to establish a general framework for constructing duality-invariant EFT observables at any given multiplicity using the double-copy. 
\section{Constructing duality-invariant Observables}\label{sec:DIObs}

Considering the decomposition in \eqn{eq:BIfromEMf}, one might naturally wonder whether other duality-invariant EFT observables can be expressed as a weighted sum over EMf amplitudes. We established in \eqn{eq:fieldTheoryPreserving} that YM building blocks satisfy BCJ relations, which is required for the KLT kernel construction. The double-copy construction of EMf, \eqn{eq:EMfDC}, then necessitates that double-copying YM with any linear combination of YMS must be duality-invariant. One path forward then is to compose YMS amplitudes with flavor-preserving HD kinematics. However, we need to be careful when arbitrarily multiplying by HD factors of $s_{ab}$, as these could inject nonlocal higher-spin modes into the amplitudes. 

Let us take a moment to explain what we mean by nonlocal in this context. Consider a four-point YMS partial amplitude generated by the Lagrangian of \eqn{eq:ymsLag}, with leg ordering $\sigma=(1234)$,
\begin{equation}
A^{\text{YMS}}_{(12)(34)} = \frac{s_{13}}{s_{12}}\equiv \diskFourpoint{1}{2}{3}{4}\,.
\end{equation}
Motivated by the construction in \eqn{eq:YMStoNLSM}, we might consider multiplying $A^{\text{YMS}}_{(12)(34)}$ by additional factors of scalar kinematics to capture HD behavior in a way that is consistent with $(12)(34)$ flavor symmetry. Under exchange of $1\leftrightarrow 2$ and $3\leftrightarrow 4$, one maps $s_{12}\leftrightarrow s_{12}$ and $s_{23}\leftrightarrow s_{13}$. Thus, we can define a HD generalization of $A^{\text{YMS}}_{(12)(34)} $ for particular values of $x$ and $y$ as follows,
\begin{equation}\label{eq:YMSHD}
A^{\text{HD}}_{(x,y)} \equiv s_{12}^x (s_{13}s_{23})^yA^{\text{YMS}}_{(12)(34)}\,.
\end{equation}
Since the kinematic prefactors that multiply $A^{\text{YMS}}_{(12)(34)}$ are functional dressing on the flavor-pairs, $A^{\text{HD}}_{(x,y)}$ by definition still satisfies BCJ relations. Thus, double-copy construction of $A^{\text{HD}}_{(x,y)}$ with YM will still yield a duality-invariant amplitude for any values of $x$ and $y$. 

However, not all values of $x$ and $y$ will yield S-matrix elements generated by a local quantum theory. Consider the case where $x=0$ and $y=1$. Evaluating $A^{\text{HD}}_{(0,1)}$ in the limit where $s_{12}\rightarrow 0$, the amplitude factorizes into a product of lower-point functions, summed over internal particle states, $h$,
\begin{equation}
\begin{split}
\lim_{s_{12} \to 0} s_{12 }\,A^{\text{HD}}_{(0,1)} = \sum_{h} A_{(12l^h)}A_{(-l^{\bar{h}}34)}= - s_{13}^3\,,
\end{split}
\end{equation} 
where the internal momentum is taken on-shell, $l^2=0$. By considering massless three-point kinematics and little-group scaling, we find this unitarity constraint has no solution when the internal particle has spin $\leq 2$, suggesting that the state space needed to produce $A^{\text{HD}}_{(0,1)}$ has a particle with $\text{spin}>2$. This is in violation with Weinberg's no-go theorem for higher-spin modes in the S-matrix \cite{Weinberg:1964ew}. These are the types of nonlocal higher-spin modes that we want to avoid when constructing duality-invariant observables with our on-shell framework. 

Thus, to guarantee that our composition of building blocks is manifestly local, we will target valid expressions with a generalization of the dimensional-reduction in \eqn{eq:genDim}. We achieve this by defining $(2d+2M)$-dimensional polarizations for our $D$-dimensional parent amplitude as follows,
\begin{equation}\label{eq:genDim2}
\begin{split}
\mathcal{E}^\varphi_a &= (\vec{0}, \epsilon_a^\varphi, \vec{0} ) \,,
\\
\mathcal{E}^\pi_a &= f_\pi^{-1}(k_a^\mu, \vec{0}, i k_a^\mu ) \,,
\end{split}
\end{equation}
where the pion decay width, $f_\pi$, is a dimensionful parameter carrying units of mass, and $\epsilon_a^\varphi$ are $2M$-dimensional orthogonal unit vectors. Furthermore, we restrict all momenta to live in the orthogonal $d$-dimensional subspace,
\begin{equation}
 \label{eq:dimRed4}\mathcal{K}_a = (k^\mu_a,\vec{0},\vec{0})\,.
\end{equation}
As we will show, amplitudes generated by plugging in these polarizations fit nicely into the framework of composing observables with $\Delta^{(\rho)}_{(\sigma)}$ building blocks. Moreover, by only considering compositions of  $\Delta^{(\rho)}_{(\sigma)}$ that are consistent with dimensional-reduction of the parent vector theory, we guarantee that the resulting double-copy amplitude is local.

\subsection{Photons with $U(1)^M$ charge} Thus far, we have established that BI amplitudes can be constructed by composing particular building blocks in the RABD. Before generalizing this construction, in this section we demonstrate how these building blocks can be mapped to a basis of complex scalar amplitudes by exploiting the polarizations defined in \eqn{eq:genDim2}. This transformation to a complex scalar basis will tease out a subtle conjectured relationship between the YMS building blocks and the scalar sector of $\mathcal{N}=4$ super Yang-Mills (sYM); suggesting that NLSM pions could inherit on-shell supersymmetry via the decomposition of \eqn{eq:YMStoNLSM}. 

To begin, let us consider a simple example with a single complex scalar. Since polarizations only appear as factors of $(\epsilon_a k_b)$ in diagrams with unpaired boundary points, a diagram that contains $\epsilon_a$ must vanish when the polarization is projected along a $\varphi$-direction, $\epsilon_a \rightarrow \mathcal{E}^\varphi_a$. When $\mathcal{E}^\varphi$ lives in $D=2d+2$ dimensions, we can give the scalars a complex structure:
\begin{equation}
\mathcal{E}^\varphi = \frac{1}{\sqrt{2}}(\vec{0}, 1,i, \vec{0} )
\qquad\mathcal{E}^{\bar{\varphi}} = \frac{1}{\sqrt{2}}(\vec{0}, 1,-i, \vec{0} )\,.
\end{equation}
Plugging these into a four-point amplitude with ordering $\sigma=(1234)$, we obtain the following set of complex YMS amplitudes in terms of our building blocks:
\begin{equation}\label{eq:cA}
\begin{split}
A_{({\bar{\varphi}} {\bar{\varphi}} {\varphi} {\varphi} )} &=\diskFourpoint{1}{4}{2}{3}+\diskFourpoint{1}{3}{2}{4}\,,
\\
A_{(\bar{\varphi} \varphi \bar{\varphi}\varphi )} &=\diskFourpoint{1}{2}{3}{4}+\diskFourpoint{1}{4}{2}{3}\,,
\\
A_{(\bar{\varphi} \varphi \varphi \bar{\varphi} )} &= \diskFourpoint{1}{3}{2}{4}+\diskFourpoint{1}{2}{3}{4}\,.
\end{split}
\end{equation}
More generally, any configuration with $k$-complex scalars of the same flavor can be reexpressed in terms of sum over $k!$ different $\Delta^{(\rho)}_{(\sigma)}$ elements. This is essentially required by Bose-symmetry. 

By inverting relations like \eqn{eq:cA} that express a triplet of complex scalar amplitudes in terms of three building blocks, $n$-point pure-scalar building blocks can be rewritten completely in terms of $U(1)^M$ scalar amplitudes, as long as $M$ is suitably large. Take for example the building block, $\Delta^{(12)(34)}_{(1234)}$, reexpressed below,
\begin{equation}\label{eq:Ctoflavor}
\diskFourpoint{1}{2}{3}{4}=\frac{1}{2}\left(A_{(\bar{\varphi} \varphi \bar{\varphi}\varphi )} +A_{(\bar{\varphi} \varphi \varphi \bar{\varphi} )}-A_{({\bar{\varphi}} {\bar{\varphi}} {\varphi} {\varphi} )}\right)\,.
\end{equation}
Naively, one would expect ``suitably large" for $n$-point amplitudes to mean $M\geq n/4$, since one could always represent subsets of four-scalars with two-flavor lines, as was done in \eqn{eq:Ctoflavor}. However, this may be a little too conservative for high-multiplicity amplitudes.

Consider the case of $M=3$ complex scalars, which describes the bosonic sector of $\mathcal{N}=4$ sYM. In principle, one could express any $\Delta^{(\rho)}_{(\sigma)}$ completely in terms of $\mathcal{N}=4$ sYM scalar amplitudes at least up to 12-point without any difficulty. However, as we alluded to in \sect{sec:AmpRel}, as we climb to higher multiplicity, an increasing number of the pure-scalar building blocks will vanish as they create disconnected areas on the interior of the graph, due to the Feynman rules of \eqn{eq:ymsLag}. An example of this behavior was shown in \eqn{eq:sixteenPointRel}. 

Considering this feature, one could imagine there is a bounded value for $M$ at which all scalar building blocks of Yang-Mills are spanned by the amplitudes in a $U(1)^M$ YMS theory. In general, the growth of nonvanishing diagrams is limited by the maximum multiplicity of the $2n$-point scalar contact, generated by $F^n$-type operators in the parent gauge theory, i.e., diagrams with contacts above $2n$-point must vanish. 

Thus, it is reasonable to conjecture that all building blocks from a dimension-$2n$ gauge theory, should be spanned by the dimensionally reduced $U(1)^M$ scalar theory, with $M=n+1$ -- just enough to probe the first vanishing contact at $2(n+1)$-point. For the building blocks of a dimension-four Lagrangian like Yang-Mills, this would indicate we need $M=3$ complex scalar amplitudes, like those in $\mathcal{N}=4$ sYM. Therefore, we postulate the following relationship between YMS building blocks and pure-scalar amplitudes in $\mathcal{N}=4$ sYM:
\begin{equation}\label{eq:YMSfromSUSY}
\Big[\!\!\Big[A^{\text{YMS}}_{(\sigma|\rho)}\Big]\!\!\Big]= \sum_\beta \pi_{(\rho |\beta)} A^{\mathcal{N}=4}_{(\sigma|\beta)}\,,
\end{equation}
where $\Big[\!\!\Big[A^{\text{YMS}}_{(\sigma|\rho)}\Big]\!\!\Big]$ is an equivalence class of amplitudes defined as,
\begin{equation}
\Big[\!\!\Big[A^{\text{YMS}}_{(\sigma|\rho)}\Big]\!\!\Big]\equiv\Big\{ A^{\text{YMS}}_{(\sigma|\rho)}+ \sum_{\tau}\Delta^{(\tau)}_{(\sigma)}\Big|\,\Delta^{(\tau)}_{(\sigma)}\in \crustlessSet \Big\}\,,
\end{equation}
and $\crustlessSetinLine$ denotes the set of crustless diagrams -- those that vanish in the RABD of Yang-Mills. In the above expression $\sigma$ is the color ordering, and matrix elements $\pi_{(\rho|\beta)}$ are indexed by the possible flavor structures of each theory, $\rho$ and $\beta$. We have verified that \eqn{eq:YMSfromSUSY} is indeed true for a subset of our building blocks through 18-point. 

An all-multiplicity proof of this conjecture, likely by way of a solution to the pizza crust problem, could have intriguing consequences -- as it would allow us to express all the pure-scalar amplitudes in this paper as sums over $\mathcal{N}=4$ sYM scalar amplitudes. Furthermore, since all the relations derived in this paper are linear and leave external leg ordering untouched, realizing \eqn{eq:YMSfromSUSY} could allow one to imprint supersymmetry on a broad class of effective observables.
\subsection{Photons from higher-derivatives:} Now we are prepared to construct observables for duality-invariant EFTs in generality. First, we take the polarizations and momenta defined in \eqn{eq:genDim2} and \eqn{eq:dimRed4}, respectively, and state the result of their dot products:
\begin{equation}
\begin{split}
 \mathcal{E}^\varphi_{a} \mathcal{E}^\varphi_{b}&=\epsilon^\varphi_a\epsilon^\varphi_b\quad \mathcal{E}^\pi_{a} \mathcal{E}^\varphi_{b} =0 \quad\mathcal{E}^\pi_{a} \mathcal{E}^\pi_{b} =0\,,
\\\\
&\mathcal{K}_{a} \mathcal{E}^\varphi_{b}=0\quad \mathcal{K}_{a} \mathcal{E}^\pi_{b} =s_{ab} \,.
\end{split}
\end{equation}
For simplicity, we will take all the $\varphi$ scalars to live in co-orthogonal dimensions, where flavor groupings and complex structure could be reinstated by summing over a suitable set of building blocks, as was done in \eqn{eq:Ctoflavor}. 

When plugging these in, different choices of dimensional-reduction will select distinct building blocks, $\Delta^{(\rho)}_{(\sigma)}$, and weight them by appropriate factors of $s_{ab}$ consistent with locality. One such choice of polarizations defined in $D=(2d+4)$-dimensions, with $(12)(47)$ legs in the space spanned by $ \mathcal{E}^\varphi$ polarizations and the remaining $(3568)$ legs projected along pion directions, yields the following:
\begin{align}
A^{(12)(47)_\varphi (3568)_\pi}_{(12345678)} &= \diskEightpointAB{1}{2}{}{}{4}{7}{}{}\Bigg|_{\epsilon\rightarrow k}
\\
&= \diskEightpointhAB{1}{2}{3}{5}{4}{7}{6}{8}+\diskEightpointhAB{1}{2}{3}{8}{4}{7}{5}{6}+\diskEightpointhAB{1}{2}{3}{6}{4}{7}{5}{8}
\\
&= \diskEightpointhAB{1}{2}{3}{5}{4}{7}{6}{8}+\diskEightpointhAB{1}{2}{3}{8}{4}{7}{5}{6}\label{eq:newAmp8}\,.
\end{align}
Since $A^{(12)(47)_\varphi (3568)_\pi}_{(12345678)}$ is simply a linear combination of YMS amplitudes, akin to the construction of NLSM amplitudes in \eqn{eq:YMStoNLSM}, the result of double-copying this with Yang-Mills must also yield a duality-invariant observable. This motivates a general expression for duality-invariant building blocks, $M^{(\rho)}_{(\sigma)}$, that is consistent with field theoretic locality, in terms of YM building blocks appearing in the RABD,
\begin{equation}\label{eq:genBlock}
M^{(\rho)}_{(\sigma)} \equiv \Delta_{(\sigma)}^{(\rho)}\Big|_{\epsilon\rightarrow k}\,.
\end{equation}
The mass-dimension of these $M^{(\rho)}_{(\sigma)}$ relative to YMS is simply $\big[M^{(\rho)}_{(\sigma)}\big] = |\sigma|-|\rho|$. By applying the Ward identity of \eqn{eq:GIrelB},  it follows that $M^{(\rho)}_{(\sigma)}$ can be expressed completely in terms of YMS amplitudes. Thus, the matrix elements constructed from these building blocks via the double copy, $\mathcal{M}^{\text{EMf-EFT}}_{(\rho)}$, will acquire duality invariance via a sort of ``kinematic-projection" from the space of EMf amplitudes:
\begin{align}
\mathcal{M}^{\text{EMf-EFT}}_{(\rho)} &\equiv A^{\text{YM}}_{(\sigma)}S_{[\sigma|\tau]} M^{(\rho)}_{(\tau)}=\sum_{\beta} s_{(\beta)} \mathcal{M}^{\text{EMf}}_{(\rho\cup \beta)}\,.
\end{align}
The last equality is a result of expressing $M^{(\rho)}_{(\tau)}$ as a sum over factors of $s_{(\beta)}A^{\text{YMS}}_{(\tau|\rho\cup \beta)}$, up to an overall sign. We note that since $\Delta^{(\varnothing)}_{(\sigma)}=0$ for the YM building blocks needed in this construction, Born-Infeld amplitudes are the highest mass-dimension duality-invariant observables built out of $M^{(\rho)}_{(\sigma)}$. 

In principle, there could be additional operators above BI mass dimension that are constructed via pole-cancelling kinematic factors, in a similar fashion to what was achieved at four-point in \cite{Carrasco:2019yyn}. In this sense, $\mathcal{M}^{\text{EMf-EFT}}_{(\rho)}$ should be seen as the set of building blocks at the lowest rung in duality-invariant EFT, analogous to how $\text{YM}$ and $(DF)^2+\text{YM}$ \cite{Johansson:2017srf} observables are the leading order terms in the low energy effective action of the open super- and bosonic-string \cite{Broedel:2013tta, Huang:2016tag}, respectively.

Before concluding, we briefly examine one of the novel matrix elements nested in $M^{(\rho)}_{(\sigma)}$ that lies between YMS and NLSM mass dimension. Consider the following cut on the last term in the amplitude of \eqn{eq:newAmp8},
\begin{equation}
\diskEightpointhABig{1}{2}{3}{8}{4}{7}{5}{6}\,\Bigg|^{s_{12}\text{-cut}}_{s_{4567}\text{-cut}}\,=\,\eightPointCut{}\,.
\end{equation}
Let us focus on the four-point factor in the center of the cut, which we will call $A^{\chi}_{(\sigma)}$, with on-shell legs, $\sigma=(k_{12},k_3,k_{4567},k_8)$. We can express a particular helicity configuration of $A^{\chi}_{(1234)}$ in terms of a scalar-vector YMS amplitude -- using the conventions of \cite{Dixon:1996wi} -- as follows:
\begin{equation}\label{eq:newHD4point}
A^{\chi}_{ (g^-,\pi,g^+,\pi)} = s_{24} A^{\text{YMS}}_{ (g^-,\pi,g^+,\pi)} = f_\pi^{-2}\frac{s_{13}\langle1 | 2|3 ]^2}{s_{12}s_{23}}\,.
\end{equation}
Since this amplitude still satisfies BCJ relations, it can be used to construct higher-derivative EMf amplitudes through double-copy construction with Yang-Mills,
\begin{equation}
\begin{split}
\mathcal{M}^{\text{HD}}_{(h^{-},\gamma^{-},h^+,\gamma^+)} & \equiv A^{\text{YM}}_{ (g^-, g^-,g^+, g^+)}\otimes A^{\chi}_{ (g^-,\pi,g^+,\pi)} 
\\
&=s_{24}\mathcal{M}^{\text{EMf}}_{(h^{-},\gamma^{-},h^+,\gamma^+)}
\\
&= \frac{\langle1 | 2|3 ]^2 \langle 12\rangle^2 [34]^2}{s_{12}s_{23}} 
\\
&= \frac{\langle1 | 2|3 ]}{\langle3 | 2|1 ] }\langle 12\rangle^2 [34]^2\,.
\end{split}
\end{equation}
Evidently, $\mathcal{M}^{\text{HD}}$ is a polynomial function of kinematics, up to an overall phase that controls the little group scaling. Thus, it is a pure four-point contact. Gravitational chiral operators that generate this matrix element have been studied in \cite{Bossard:2011tq}. It would be interesting if these operators have a similar dimensional-reduction description on the gravity side, as $A^{\chi}_{(\sigma)}$ has on the gauge theory side. 

Let us point out that it is somewhat surprising that two amplitudes with nonvanishing singularity structure, $A^{\text{YM}}$ and $A^\chi$, double-copy to a pure contact term absent of singularities. There is a nice explanation for this provided by the three-point functions of each theory. By performing an exhaustive search, we find that the only nonvanishing on-shell three-point interaction consistent with factorization is the following $D$-dimensional vertex,
\begin{equation}
A^{\chi}_{ (g, g,\pi )} = f_\pi^{-1} \kappa_{3}^{(1)}\kappa_{3}^{(2)}\,.
\end{equation}
This vertex is generated by a $\text{tr}(\pi F^2)$ interaction, which is \textit{symmetric} under color ordering. In contrast, the color ordered YM vertex is anti-symmetric. Thus, the double-copy to gravity, $\mathcal{M}^{\text{HD}}_3=A^{\text{YM}}_3A^{\chi}_3$, must vanish in concordance with Bose symmetry, in which case, the four-point amplitude $\mathcal{M}^{\text{HD}}_4$ should be some type of contact, as we have found.

To conclude, we note that the matrix element of \eqn{eq:newHD4point} can be constructed with the following dimensional-reduction on the full YM four-point amplitude,
\begin{align}
\mathcal{K}_{a} &= (k^\mu_{a},0,\vec{0})
\\
\mathcal{E}_{1,2}^g &= (\epsilon^\mu_{1,2},0,\vec{0})
\\
\mathcal{E}_{3,4}^\pi &= (k^\mu_{3,4},0,ik^\mu_{3,4}).
\end{align}
By expanding our library of polarizations to include gluons, one can thus compose factors of $\Delta^{(\rho)}_{(\sigma)}$ to target HD scalar-vector amplitudes that are consistent with BCJ relations, like those studied in \cite{Carrasco:2022sck}. 
\section{Discussion}\label{sec:Discussion}
In summary, we have established a framework for constructing effective observables for duality-invariant theories at all-multiplicity with the building blocks of \eqn{eq:genBlock}. We have done so by considering how the generalized dimensional-reductions in \eqn{eq:genDim} and novel Ward identities derived in \eqn{eq:GIrelA} and \eqn{eq:GIrelB} act on a basis of pure-scalar YMS amplitudes. We were able to do this by introducing a novel amplitude decomposition in \eqn{eq:pureVecRABD} (RABD) that preserves partial amplitude relations, and can be applied to any $D$-dimensional gauge theory amplitude, not just Yang-Mills. By studying the behavior of these building blocks, we derived new amplitude relations between NLSM and YMS amplitudes in \eqn{eq:YMStoNLSM}, and discovered additional higher-derivative BCJ satisfying amplitudes, \eqn{eq:newHD4point}, generated by a three-point operator,  $\text{tr}(\pi F^2)$, that is symmetric under color ordering. 

The directions of future study are vast. The obvious first step is to better understand gauge theory amplitudes in terms of these building blocks. In this study, we have just focused on pure-scalar building blocks in Yang-Mills. But given the universal nature of our decomposition, one could study them at the integrand level \cite{He:2015yua,Zhou:2022djx}, and in higher-derivative vector theories. 

Take for example a building block that vanishes in Yang-Mills, $\Delta^{(14)(25)(36)}_{(123456)}$, but is generated by the six-point contact appearing in the construction of $\text{YM}+F^3$, a theory whose on-shell behavior has been studied extensively in \cite{Carrasco:2022lbm,Broedel2012rc, He:2016iqi},
\begin{equation}
\diskSixpointStraight{1}{4}{2}{5}{3}{6}^{\text{YM}+F^3}=1\,.
\end{equation}
It would be worthwhile understanding how higher-point contacts like this augment the pizza crust problem. This would affect the decomposition of our building blocks into $U(1)^M$ complex scalar amplitudes, like in \eqn{eq:Ctoflavor}. To this end, we have conjectured that at ``sufficiently large $M$"  there should be a $U(1)^M$ scalar theory that spans all scalar building blocks at a given mass dimension. An explicit statement of this conjectured relationship was provided for dimension-four YMS amplitudes in \eqn{eq:YMSfromSUSY}. Determining whether or not this is true would have serious implications for mapping SUSY to duality invariance, in the spirit of \cite{Carrasco:2013qia}.

Furthermore, there is a large body of literature on duality-invariant operators \cite{Brace:1999zi,Hatsuda:1999ys,Chemissany:2011yv,BabaeiVelni:2016qea,BabaeiVelni:2019ptj,Bandos:2020jsw,Carrasco:2022jxn}, along with some recent studies that have enumerated the space of Einstein-Maxwell effective operators using Hilbert series methods \cite{Cano:2021tfs,Cano:2021hje}. It would be interesting to find a mapping between these Lagrangian-level operators and the matrix elements that we have constructed here.

Finally, while in this work we have focused on photon amplitudes, the double-copy construction of $\Delta^{(\rho)}_{(\sigma)}$ with itself should lead to observables of flavored scalars coupled to external gravitons -- arguably describing higher-derivative tidal effects \cite{Bern:2020uwk,Haddad:2020que,Aoude:2020ygw} with radiative corrections \cite{Shen:2018ebu,Carrasco:2020ywq} of relevance for gravitational wave physics. In addition, better understanding building blocks from HD vector theories could elucidate a relationship between EMf duality invariance and helicity selection rules for higher-dimensional gravity counterterms \cite{Broedel2012rc,He:2016iqi}, like the evanescent $R^2$ Gauss-Bonnet term \cite{Bern:2017tuc}. We find the prospect of targeting these HD gravity operators a particularly exciting direction of future work. 

\noindent \textbf{Acknowledgments:} The author would like to thank John Joseph Carrasco, Alex Edison, Lance Dixon, Matthew Lewandowski, James Mangan, Aslan Seifi, Bogdan Stoica and Suna Zekio\u{g}lu for a combination of insightful conversations, collaboration on related research, and feedback on the manuscript. The author also thanks Katheryn Scott for naming of the pizza crust problem. N.H.P. acknowledges the Northwestern University Amplitudes and Insights group, the Department of Physics and Astronomy, and Weinberg College for support.
\bibliography{Refs_U1fromDC.bib}
\end{document}